\documentclass[aps,pra,reprint,superscriptaddress,amsmath,amsfonts,amssymb,longbibliography]{revtex4-1}

\usepackage[utf8]{inputenc}
\usepackage{graphicx}
\usepackage{float}

\usepackage{xcolor}

\usepackage{amssymb}
\usepackage{dsfont}

\usepackage{physics}

\usepackage{listings}

\usepackage{color}

\usepackage{hyperref}

\usepackage[nodisplayskipstretch]{setspace}

\usepackage{hyperref}

\begin{document}

\title{The singularities of the rate function of quantum coherent work in one-dimensional transverse field Ising model}

\author{Bao-Ming Xu}
\email{xubm2018@163.com}

\affiliation{Shandong Key Laboratory of Biophysics, Institute of Biophysics, Dezhou University, Dezhou 253023, China}%

\author{Chao-Quan Wang}
\affiliation{School of Science, East China University of Technology, Nanchang 330013, China}

\date{Submitted \today}

\begin{abstract}
Quantum coherence will undoubtedly play a fundamental role in understanding of the dynamics of quantum many-body systems, thereby to reveal its genuine contribution is of great importance. In this paper, we specialize our discussions to the one-dimensional transverse field quantum Ising model initialized in the coherent Gibbs state. After quenching the strength of the transverse field, the effects of quantum coherence are studied by the rate function of quantum work distribution. We find that quantum coherence not only recovers the quantum phase transition destroyed by thermal fluctuations, but also generates some entirely new singularities both in the static state and dynamics. It can be manifested that these singularities are rooted in spin flips causing the sudden change of the domain boundaries of spin polarization. This work sheds new light on the fundamental connection between quantum critical phenomena and quantum coherence.
\end{abstract}

\maketitle
\section{Introduction}
Experiments with ultracold atom \cite{Levin2012,Yukalov2011,Bloch2008,Greiner2002} and ion-trap \cite{Porras2004,Kim2009,Jurcevic2014} provide opportunities to manipulate and control the system, both in space and in time, with an unprecedented accuracy as compared to any solid-state counterpart, preserving coherence over long time scales. With these experimental advances, an in-depth study of the dynamics of quantum many-body systems becomes possible. It aims to shed light on some fundamental issues of quantum mechanics that have been recently resurfaced, such as thermalization of closed systems \cite{Rigol2008,Ueda2020} or the emergence of universality in the dynamics across a critical point \cite{Dziarmaga2010,Nigro2019,Heyl2015}. With these achievements, it has become particularly relevant to develop specific theoretical frameworks that could work in out-of-equilibrium situations, overcoming the limitations of linear-response and perturbation theory. In this regard, an important progress of stochastic thermodynamics has been made \cite{Seifert2012,Seifert2019,Ciliberto2017,Hanggi2011,Esposito2009}, where large deviations \cite{Touchette2009,Gambassi2012} have allowed for a better understanding of the emerging phenomena both in the steady states and their fluctuations \cite{Garrahan2010,Perfetto2022,Garrahan2018}. Particularly relevant has been the discovery of the exact fluctuation theorems, which hold for the system arbitrarily far from equilibrium \cite{Seifert2012,Seifert2019,Ciliberto2017,Hanggi2011,Esposito2009}, being able to characterize the full non-linear response of the system to any (perturbative or not) driving. These rely on the analysis of thermodynamic key concepts as work, heat, and entropy, which represent stochastic variables with definite probability distributions.

One of the most intriguing examples of studying the dynamics of quantum many-body systems in this nonequilibrium thermodynamical formulation is dynamical phase transition \cite{Heyl2013,Heyl2019,Heyl2018}. There has been quite a remarkable amount of activity uncovering the features of dynamical phase transition in a range of physical models including Hermitian \cite{Heyl2013,Lang2018,Halimeh2021a,Cheraghi2023,Corps2022a,Nicola2021,Bandyopadhyay2021,Yu2021,Arenas2022,Corps2022b} and non-Hermitian \cite{Zhou2018,Zhou2021a,Mondal2022} systems, topological matter \cite{Vajna2015,Schmitt2015,Jafari2016,Jafari2017,Sedlmayr2018,Jafari2018,Zache2019,Mas2020,Okugawa2021,Sadrzadeh2021,Ding2020}, Floquet systems \cite{Sharma2014,Yang2019,Zamani2020,Zhou2021b,Jafari2021,Hamazaki2021,Zamani2022,Luan2022,Jafari2022,Naji2022}, and many-body localized systems \cite{Halimeh2019,Trapin2021,Halimeh2021b}, etc. Dynamical phase transition, manifested as real-time singularities in time-evolving quantum system after quenching a set of control parameters of its Hamiltonian, is indeed a dynamical analogue of equilibrium phase transition. It is defined via the temporal nonanalytic behaviour of the rate function, namely dynamical free energy density, an analogue of the free energy, defined for the Loschmidt echo rather than the partition function \cite{Heyl2013}. Loschmidt echo is just the probability that the quench protocol does not do work on the system, therefore, the dynamical phase transition shows up as nonanalytic behavior in the work distribution function: The cusps featured by the rate function of quantum work distribution at critical times \cite{Heyl2013}. This nonanalytic behavior can not be observed at nonzero temperatures due to the presence of thermal fluctuating \cite{Abeling2016}. All these afore mentioned studies require the initial ground or thermal equilibrium state of the system in which there is no quantum coherence. In the quantum realm, quantum coherence will undoubtedly play a fundamental role, for instance it is strictly related to the irreversible work \cite{Landi2021,Varizi2020,Varizi2021,Francica2019,Santos2019,Xu2018} and it leads a genuine quantum contribution to quantum critical phenomena \cite{Xu2018,Li2020,Mao2021,Swan2020,Lv2022,Pires2021,Wanf2022}. Very recently, a clear connection among coherent signatures, enhanced work extraction and the critical behaviors in quadratic fermionic systems has been found via the Kirkwood-Dirac quasiprobability approach \cite{Santini2023}. Therefore, reveal the effects of quantum coherence on the dynamics of quantum many-body systems are important and pivotal.

In this paper, we aim at revealing some unique effects of quantum coherence on the critical phenomena of quantum many-body system in the
nonequilibrium thermodynamical formulation with the aid of large deviation theory \cite{Touchette2009,Gambassi2012}. To accomplish this, we specialize our discussions to the quenched one-dimensional transverse field quantum Ising model, one of two prototypical models to understand the quantum phase transition \cite{Sachdev2011}. The system is first prepared in the coherent Gibbs state \cite{Xu2018,Lostaglio2015,Kwon2018}, an extreme example of coherent resources in quantum thermodynamics, and then quenched by suddenly changing the transverse field, a common way to drive an isolated quantum system out of equilibrium. By treating the quench as a thermodynamic transformation, the quantum work distribution and its rate function are determined by Gaussian measurement scheme \cite{Xu2021} rather than projective measurement, because projective measurement will completely destroy quantum coherence. With these nonequilibrium thermodynamical formulation at hand, we then discuss the effects of quantum coherence on the dynamics of quantum many-body systems, especially on some dynamical singularities.

This paper is organized as follows: In the next section, we introduce the quenched one-dimensional transverse Ising model and derive the rate function of quantum work distribution, which will be considered in the following. The results that the static and dynamical singularities of rate function are discussed in Sec. III and Sec. IV, respectively. Finally, Sec. V closes the paper with some concluding remarks.

\section{Quenched one-dimensional Transverse Field Ising model}
The Hamiltonian of the one-dimensional transverse field Ising model is
\begin{equation}\label{}
  \hat{H}\bigl(\lambda_t\bigr)=-\frac{J}{2}\sum_{j=1}^{N}\bigl[\hat{\sigma}_j^z\hat{\sigma}_{j+1}^z+\lambda\hat{\sigma}_j^x\bigr],
\end{equation}
where $J$ is longitudinal spin-spin coupling, $\lambda$ is a dimensionless parameter measuring the strength of the transverse field with respect to the longitudinal spin-spin coupling. In this work, we set $J=1$ as the overall energy scale and only consider $\lambda\geq0$ without loss of generality. $\hat{\sigma}^{\alpha}_{j}$ $(\alpha=x,y,z)$ is the spin-1/2 Pauli operator at lattice site $j$ and the periodic boundary conditions are imposed as $\hat{\sigma}^{\alpha}_{N+1}=\hat{\sigma}^{\alpha}_{1}$. Here we only consider that $N$ is even. The one-dimensional quantum Ising model is the prototypical, exactly solvable example of a quantum phase transition, with a quantum critical point at $\lambda_c=1$ separating a quantum paramagnetic phase at $\lambda>\lambda_c$ from a ferromagnetic one at $\lambda<\lambda_c$.

After Jordan-Wigner transformation and Fourier transforming, the Hamiltonian becomes a sum of two-level systems \cite{Russomanno2012}:
\begin{equation}\label{}
\hat{H}(\lambda)=\sum_k\hat{H}_k(\lambda).
\end{equation}
Each $\hat{H}_k(\lambda)$ acts on a two-dimensional Hilbert space generated by $\{\hat{c}^\dag_k\hat{c}^\dag_{-k}|0\rangle,~|0\rangle\}$, where $|0\rangle$ is the vacuum of the Jordan-Wigner fermions $\hat{c}_k$, and can be represented in that basis by a $2\times2$ matrix
\begin{equation}\label{}
 \hat{H}_k(\lambda)=(\lambda-\cos k)\hat{\sigma}^z+\sin k\hat{\sigma}^y,
\end{equation}
where $k=(2n-1)\pi/N$ with $n=1\cdots N/2$, corresponding to antiperiodic boundary conditions for $N$ is even. The instantaneous eigenvalues are $\varepsilon_{k}(\lambda)$ and $-\varepsilon_{k}(\lambda)$ with
\begin{equation}\label{}
\varepsilon_{k}(\lambda)=\sqrt{(\lambda-\cos k)^2+\sin^2k}.
\end{equation}
The corresponding eigenvectors are
\begin{equation}\label{}
|\varepsilon^+_{k}(\lambda)\rangle=
\bigl[\cos\theta_k+i\sin\theta_k\hat{c}^\dag_k\hat{c}^\dag_{-k}\bigr]|0\rangle
\end{equation}
and
\begin{equation}\label{}
|\varepsilon^-_{k}(\lambda)\rangle=\bigl[i\sin\theta_k+\cos\theta_k\hat{c}^\dag_k\hat{c}^\dag_{-k}\bigr]|0\rangle,
\end{equation}
respectively, where $\theta_k$ is determined by the relation
\begin{equation}\label{}
e^{i\theta_k}=\frac{\lambda-\varepsilon_{k}(\lambda)-e^{-ik}}{\sqrt{\sin^2k+[\lambda-\cos k-\varepsilon_{k}(\lambda)]^2}}.
\end{equation}

In order to explore the effects of quantum coherence, we consider the system to be initially in the coherent Gibbs state
\begin{equation}\label{}
  |\psi_{th}(\lambda)\rangle=\bigotimes_{k}\sqrt{\frac{e^{-\beta\hat{H}_k(\lambda)}}{Z_{k}(\lambda)}}
  \biggl(|\varepsilon^+_{k}(\lambda)\rangle+|\varepsilon^-_{k}(\lambda)\rangle\biggr)
\end{equation}
compared with the corresponding thermal equilibrium state
\begin{equation}\label{}
  \hat{\rho}_{th}(\lambda)=\bigotimes_{k}\frac{e^{-\beta\hat{H}_k(\lambda)}}{Z_{k}(\lambda)}.
\end{equation}
In the expression above, $\beta=1/T$ is the inverse of the temperature (we have set Boltzmann constant $k_B=1$), $Z_k(\lambda)=2\cosh[\beta\varepsilon_{k}(\lambda)]$ refers to the partition function of mode $k$.

To study the dynamics of quantum many-body systems, one can analyze an experimentally tractable quantity: the work done in a nonequilibrium process, such as unitary evolution, sudden quench, or the mixture of them. Without loss of generality, the nonequilibrium process can be expressed as $\mathcal{L}_t(\cdot)$, which transforms the system from its initial state $\hat{\rho}(0)$ to the final state $\hat{\rho}(t)=\mathcal{L}_t\bigl(\hat{\rho}(0)\bigr)$, changing the system Hamiltonian from $\hat{H}(\lambda)$ to $\hat{H}(\lambda')$. The work done in this process is usually determined by two-point measurement scheme: performing two projective energy measurements at the beginning and the end of external protocol. If the initial state $\hat{\rho}(0)$ has quantum coherence, the first projective measurement will destroy it and severely impact the system dynamics and the work statistics. In order to include the effects of initial quantum coherence, a Gaussian measurement scheme was proposed in Ref. \cite{Xu2021}. Based on this Gaussian measurement scheme, the cumulant generating function of the work distribution function $G(R,t)=\int dWP(W,t)e^{-RW}$ ($R\in\mathbb{R}$) can be expressed as
\begin{equation}\label{}
  G(R,t)=e^{\frac{R^2\sigma^2}{2}}\prod_kG_k(R,t)
\end{equation}
with
\begin{equation}\label{Gk}
  G_k(R,t)=\mathrm{Tr}\biggl[e^{-R\hat{H}_k(\lambda')}\mathcal{L}_t
  \Bigl(e^{\frac{R\hat{H}_k(\lambda)}{2}}\hat{\varrho}(0)e^{\frac{R\hat{H}_k(\lambda)}{2}}\Bigr)\biggr],
\end{equation}
where $\hat{\varrho}(0)$ is the system state after the first Gaussian measurement with error $\sigma$. If the system is initialized in the thermal equilibrium state, i.e., $\hat{\rho}(0)=\hat{\rho}_{th}(\lambda)$, the system state after the first Gaussian measurement is still the thermal equilibrium state, i.e., $\hat{\varrho}(0)=\hat{\rho}_{th}(\lambda)$. If the system is initialized in the coherent Gibbs state, i.e., $\hat{\rho}(0)=|\psi_{th}(\lambda)\rangle\langle \psi_{th}(\lambda)|$, the system state after the first Gaussian measurement is $\hat{\varrho}(0)=\otimes_k\hat{\varrho}_k(\lambda)$ with
\begin{equation}\label{}
\hat{\varrho}_k(\lambda)=\frac{1}{Z_k(\lambda)}
\begin{pmatrix}
e^{-\beta \varepsilon_{k}(\lambda)} & e^{-\bigl(\frac{\varepsilon_{k}(\lambda)}{\sigma}\bigr)^2} \\
e^{-\bigl(\frac{\varepsilon_{k}(\lambda)}{\sigma}\bigr)^2} & e^{\beta \varepsilon_{k}(\lambda)}
\end{pmatrix}.
\end{equation}

The work distribution obeys a large deviation form $P(W,t)\sim e^{-Nr(w,t)}$ with a rate function $r(w,t)\geq0$ depending on the work density $w=W/N$. In the thermodynamic limit one can derive an exact result for $r(w,t)$: According to the Ga\"{a}rtner-Ellis theorem \cite{Touchette2009} it is just the Legendre transform
\begin{equation}\label{rw}
  -r(w,t)=\inf_{R\in\mathbb{R}}[wR-c(R,t)],
\end{equation}
where
\begin{equation}\label{}
  c(R,t)=-\lim_{N\rightarrow\infty}\frac{1}{N}\ln G(R,t)
\end{equation}
is the rate function for the cumulant generating function of the work distribution function $G(R,t)$. $c(R,t)$ is always concave and continuous inside the relevant domain of definition. Because $G(R,t)$ splits into products over $k$, the function $c(R,t)$ can be simplified significantly:
\begin{equation}\label{cr}
\begin{split}
  c(R,t)=&-\lim_{N\rightarrow\infty}\frac{1}{N}\sum_k\ln G_k(R,t) \\
        =&-\frac{1}{\pi}\int_0^\pi dk\ln G_k(R,t),
\end{split}
\end{equation}
here, the constant term $\frac{R^2\sigma^2}{2 N}$ is disappeared.

\section{Static singularity of rate function}
In this section, we investigate the effects of quantum coherence on the properties of quantum work performed by the quench of transverse field from $\lambda$ to $\lambda'$. For this purpose, the system does not need to evolve after quench because it has no influence on work statistics. In this sense, we believe the system state is static and denote the singularities discussed in the following as the static singularities. The cumulant generating function (\ref{Gk}) can be expressed as
\begin{equation}\label{SGk}
  G_k(R)=\mathrm{Tr}\Bigl[e^{-R\hat{H}_k(\lambda')}e^{\frac{R\hat{H}_k(\lambda)}{2}}\hat{\varrho}_k(\lambda)e^{\frac{R\hat{H}_k(\lambda)}{2}}\Bigr].
\end{equation}
Substituting Eq. (\ref{SGk}) into Eq. (\ref{cr}) and Eq. (\ref{rw}), one can obtain the rate function $r(w)$.

\begin{figure}
\begin{center}
\includegraphics[width=8cm]{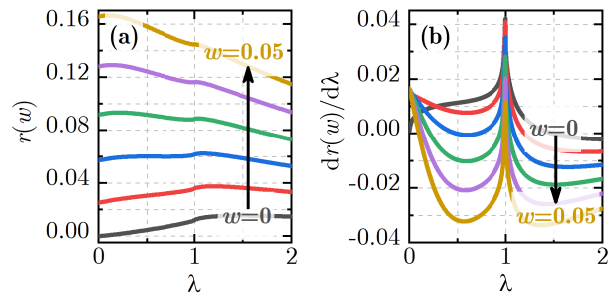}
\caption{(Color online) The curves of (a) rate function $r(w)$ and (b) its susceptibility $dr(w)/d\lambda$ as the functions of $\lambda$ for $w=0, 0.01, 0.02, 0.03, 0.04$ and $0.05$ at very low temperature $T=0.01$. The intensity of the quench is $0.01$, i.e., $\lambda'=\lambda+0.01$. It should be noted that the measurement error $\sigma$ has no effect because there is no quantum coherence at very low temperature.}
\label{figure1}
\end{center}
\end{figure}

In the low-temperature limit, the occupation number of only ground state energy is significant, and there is no quantum coherence. After quenching, the system is no longer in the ground state, but a superposition of different energy levels, causing the expectation value of work to be shifted. In Fig. \ref{figure1}(a), we plot rate function $r(w)$ for a series of sudden quenches with amplitude $|\lambda'-\lambda|=0.01$. The rate function changes smoothly with transverse field $\lambda$. The interpretation of this smooth behaviour is straight: Continues to increase the transverse field, the energy levels of the system changes smoothly. It is well known that the increasing transverse field drives the Ising model from ferromagnetically ordered phase at $\lambda<1$ to a paramagnet at $\lambda>1$. This quantum phase transition is described by the nonanalytic behavior of the susceptibilities about magnetization \cite{Vojta2003,Sachdev2011} and fidelity \cite{Gu2008,Albuquerque2010,You2007,Chen2008,Gu2010,Cozzini2007a,Cozzini2007b,Chen2007,Abasto2008} at critical point $\lambda=1$. Herein, the quantum phase transition in the Ising model is also observed, where the nonanalytic behavior of the susceptibility of the rate function occurs at the critical point $\lambda=1$ [see Fig. \ref{figure1}(b)]. This can be understood as follows: Work is performed to drive the system across the critical region and, due to the vanishing energy gap, the system becomes extremely susceptible to the driving, thereby sharpening the susceptibility of the rate function of work distribution.

\begin{figure}
\begin{center}
\includegraphics[width=8cm]{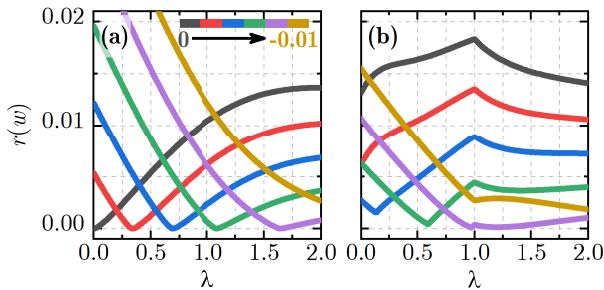}
\caption{ (Color online) The curves of rate function $r(w,t)$ as the function of $\lambda$ for $w=0, -0.002, -0.004, -0.006, -0.008$ and $-0.01$ at temperature $T=1$. The system is initially in (a) thermal equilibrium and (b) coherent Gibbs states. The intensity of the quench is $0.01$, i.e., $\lambda'=\lambda+0.01$. The measurement error is $\sigma=1$.}
\label{figure2}
\end{center}
\end{figure}

With temperature increasing, the quantum phase transition is destroyed, thereby no point of nonanalyticity can be observed in the rate function and its susceptibility [see Fig. \ref{figure2}(a)]. The temperature increases the occupation number of excited states, and the quantum coherence between energy levels becomes significant in coherent Gibbs state. Taking coherent Gibbs state as the initial state of the system, we plot the rate function $r(w)$ in Fig. \ref{figure2}(b). The rate function shows a kink at critical point $\lambda=1$, that is to say, with the aid of quantum coherence, the disappeared nonanalytic behavior in the susceptibility of the rate function is replaced by a kink singularity in the rate function itself. In this case, the kink singularity induced by quantum coherence is the first order, a new form of the demonstration of quantum phase transition in Ising model. In addition to the kink at critical point $\lambda=1$, some new kinks for different work densities can be found at the ferromagnetic region $\lambda<1$ [see the blue, green and pink curves in Fig. \ref{figure2}(b)], which implies an entirely new kind of singularity beyond the traditional quantum phase transitions.

Before explaining the singularities caused by quantum coherence, let us briefly review quantum phase transition at ground state, see Refs. \cite{Dziarmaga2005,Cincio2007}. Quantum phase transition, driven by quantum fluctuations, takes place at the critical value $\lambda = 1$ of the transverse field. When $\lambda\gg1$, the ground state is a paramagnet $|\rightarrow\rightarrow\rightarrow\cdots\rightarrow\rangle$ with all spins polarized along the $x$-axis. On the other hand, when $\lambda\ll1$, then there are two degenerate ferromagnetic ground states with all spins pointing either up or down along the $z$-axis: $|\uparrow\uparrow\uparrow\cdots\uparrow\rangle$ or $|\downarrow\downarrow\downarrow\cdots\downarrow\rangle$. However, when $N\rightarrow\infty$, then energy gap at $\lambda = 1$ tends to zero (quantum version of the critical slowing down) and it is impossible to pass the critical point at a finite speed without exciting the system. As a result, the system ends in a quantum superposition of states like $|\cdots\uparrow\downarrow\downarrow\downarrow\downarrow\downarrow\uparrow\uparrow\uparrow\uparrow\uparrow\uparrow\uparrow\downarrow\downarrow
\downarrow\downarrow\uparrow\uparrow\uparrow\uparrow\uparrow\uparrow\downarrow\cdots\rangle$ with finite domains of spins pointing up or down and separated by kinks where the polarization of spins changes its orientation. Beyond the ground state, the character of the excitations also undergoes a qualitative change across the quantum critical point, which is described in the Landau quasiparticle scheme, i.e., as superpositions of nearly independent particle-like excitations \cite{Sachdev2011}.

With temperature increasing, quantum fluctuations are overwhelmed by thermal fluctuations, compelling the symmetry-breaking order-disorder transition caused by quantum fluctuations to be disappeared. Quantum coherence, as a kind of information, is used to do work and, due to its extreme sensitivity nature to energy level structure, the resulting coherent work, i.e., the work involving quantum coherence, is highly susceptible to the change of energy level structure. On the other hand, quantum coherence enhances quantum fluctuations, enabling them to overcome thermal fluctuations, thus reviving the symmetry-breaking order-disorder transition. This symmetry breaking causes the rate function of work distribution to be singular at the phase transition point. The kinks in the ferromagnetic region comes from the spin flips driven by transverse field in different energy levels. The spin flips in different energy levels generate different energy changes corresponding to different work densities. After quenching transverse field, the domain boundaries of spins pointing up or down in some energy levels to be changed suddenly, thereby occurring the kinks in the corresponding work densities. In the paramagnetic region, the system is already disordered, the effect of quantum coherence is no longer significant.

\begin{figure}
\begin{center}
\includegraphics[width=8cm]{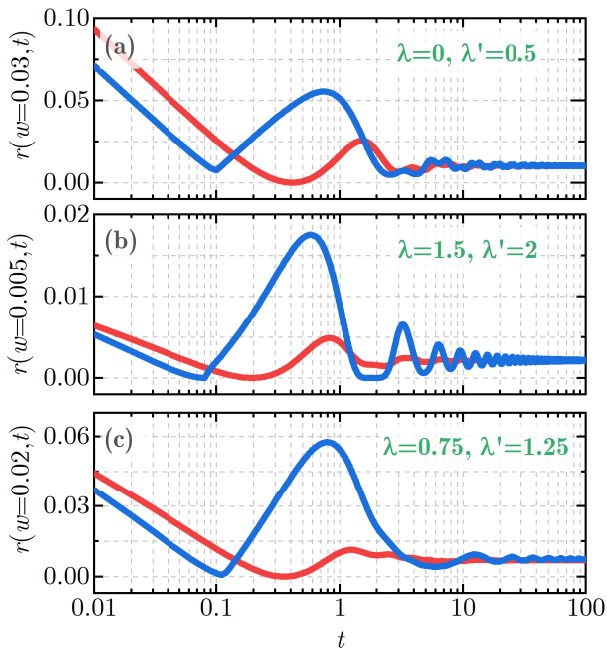}
\caption{(Color online) The dynamics of rate function $r(w,t)$ for coherent Gibbs (blue) and thermal equilibrium (red) states with temperature $T=1$. The double quenches are performed in (a) ferromagnetic region ($\lambda=0$ and $\lambda'=0.5$) and (b) paramagnetic phase region ($\lambda=1.5$ and $\lambda'=2$) and (c) across from ferromagnetic phase region to paramagnetic phase one ($\lambda=0.75$ and $\lambda'=1.25$). It should be noted that three double quenches considered have the same perturbation strength, i.e., $\delta_\lambda=\lambda'-\lambda=0.5$. In the dynamics of rate function $r(w,t)$, the work densities are given as $w=0.03, 0.02$ and $0.005$ in (a), (b) and (c), respectively. The measurement error is $\sigma=1$.}
\label{figure3}
\end{center}
\end{figure}

\section{Dynamical singularity of rate function}
In this section, we investigate the quench dynamics in the nonequilibrium thermodynamical formulation. For this, we consider a double quench protocol: At $t=0$ the parameter $\lambda$ is quenched from $\lambda$ to $\lambda'$ such that the Hamiltonian $\hat{H}(\lambda')$ governs the nontrivial time evolution $\hat{U}(t)=e^{-i\hat{H}(\lambda')t}$; at time $t$ the system is quenched back. For this double quench protocol, the cumulant generating function (\ref{Gk}) can be expressed as
\begin{equation}\label{gGk}
  G_k(R,t)=\mathrm{Tr}\Bigl[e^{-R\hat{H}_k(\lambda)}
  \hat{U}(t)e^{\frac{R\hat{H}_k(\lambda)}{2}}\hat{\varrho}_k(\lambda)e^{\frac{R\hat{H}_k(\lambda)}{2}}\hat{U}^\dag(t)\Bigr].
\end{equation}
Substituting Eq. (\ref{gGk}) into Eq. (\ref{cr}) and Eq. (\ref{rw}), one can obtain the rate function $r(w,t)$. It is well known that if the system is initialized in the ground state, the temporal nonanalytic behaviours of the rate function, featuring the dynamical phase transition, can be observed \cite{Heyl2013}. However, if the system is initially in the thermal equilibrium state at finite temperature, thermal fluctuations will destroy dynamical phase transition and the nonanalyticities will disappear \cite{Abeling2016}. Herein we consider the effects of quantum coherence on the quench dynamics by initializing the system in the coherent Gibbs state.

\begin{figure}
\begin{center}
\includegraphics[width=7.2cm]{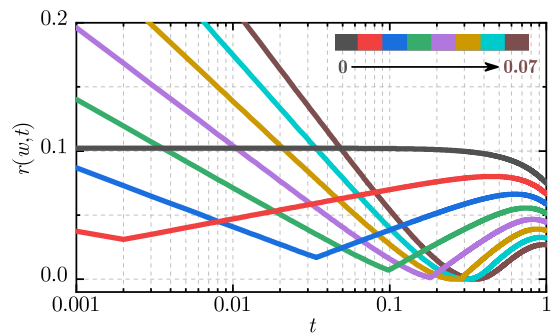}
\caption{(Color online) The short time dynamics of rate function $r(w,t)$ for $w=0, 0.01, 0.02, 0.03, 0.04, 0.05, 0.06$ and $0.07$. The double quenches are performed in ferromagnetic phase region with $\lambda=0$ and $\lambda'=0.5$. The initial state of the system is coherent Gibbs state with temperature $T=1$. The measurement error is $\sigma=1$.}
\label{figure4}
\end{center}
\end{figure}

Fig. \ref{figure3} shows the dynamics of rat function. At short time, the rate functions for both thermal equilibrium and coherent Gibbs states satisfy $r(w,t)\sim-\alpha\ln t$, which implies that work distribution $P(W,t)\sim e^{-Nr(w,t)}\sim t^{N\alpha}$, i.e., the dynamics of work distribution at short time obeys the power law. After a period of evolution, the effects of quantum coherence become significant, causing a sudden change in the dynamics of the work distribution, which otherwise changes smoothly in the absence of coherence. The sudden change of the dynamics of the work distribution is shown by the kink in the rate function [see the blue curves in Fig. \ref{figure3}]. This sudden change behavior depends on work density. Taking the ferromagnetic phase ($\lambda=0$ and $\lambda'=0.5$) as an example, we plot the short time dynamics of rate function for different work densities in Fig. \ref{figure4}. As shown in Fig. \ref{figure4}, the dynamical kinks only occur for the low but not zero work density $0<w\lesssim0.05$. With the work density increasing, the transition time required for singularity occurrence increases, but must be kept in the short region $t\lesssim0.1$. Quantum coherence in the initial state will be destroyed after sufficiently long time relaxation, even though the relaxation process is unitary. Therefore, the rate function for any work density stabilizes at a constant value. Constant rate function in the long time limit means that the work distribution and the average work no longer change over time, i.e., $\lim_{t\rightarrow0}dP(W,t)/dt=0$ and $\lim_{t\rightarrow0}d\langle W(t)\rangle/dt=0$.

These results also root in spin flips. After the first quench, the system begins to evolve with the spin flips driven by Hamiltonian $H(\lambda')$. The effect of quantum coherence is only visible when a large number of spin flips are involved. At short time, the spins have no enough time to slip and quantum coherence does not have a significant effect, therefore work distributions both obey the same power law with or without coherence. After a period of evolution, enough spins have flipped that the domain boundaries of spins pointing up or down in energy levels can be changed suddenly. These sudden changes will be smoothed by thermal fluctuations, thereby smoothing the dynamics of rate function, if the system is initially in the thermal equilibrium state. However, quantum coherence highlights the contribution of some energy levels, making their domains particularly important. The sudden change in these domains alters the work distribution so that a kink is observed in the rate function. In the long time limit, all spins are flipped and the system is disordered, eliminating the effects of quantum coherence.

\section{Conclusions}
To summarize, the effects of quantum coherence on the nonequilibrium dynamics have been studied by focusing the rate function of quantum work distribution after quenching the strength of transverse field in the one-dimensional quantum Ising model, where the quantum Ising model is initialized in the coherent Gibbs state. After quench, the rate function of quantum work distribution shows a second-order quantum phase transition at very low temperature, but it is destroyed by thermal fluctuations when temperature goes up. With the help of quantum coherence, this disappeared quantum phase transition can be recovered by a kink singularity, beyond this, another entirely new kink singularity also occurs in ferromagnetic region. The unitary time evolution of a quantum system after a sudden global quench plays a substantial role in understanding nonequilibrium quantum physics. It has been demonstrated that the initial quantum coherence plays a role only at short periods of evolution because it will be destroyed after sufficiently long time relaxation. At short time, the dynamics of work distribution obeys a power law, but can be changed suddenly with the influence of quantum coherence, causing a dynamical kink. These unique kink singularities generated by quantum coherence have been interpreted by spin flips. These unique effects of quantum coherence on work statistics shed light on the important tasks of studying what thermodynamic quantity mean in quantum mechanics and how to extend the principles of thermodynamics to the quantum regime.

\section*{acknowledgement}
This work was supported by the National Natural Science Foundation of China (Grants No. 11705099 and No. 12164003) and the Talent Introduction Project of Dezhou University of China (Grant No. 30101437).

%
\end{document}